\documentclass[aps,prl,twocolumn,superscriptaddress,notitlepage,nobibnotes]{revtex4-1}
\usepackage[colorlinks=true, citecolor=magenta, linkcolor=blue, urlcolor=blue]{hyperref}
\usepackage{graphicx}
\usepackage{amsmath}
\usepackage{amssymb}
\usepackage{changes}
\usepackage{amsfonts}
\usepackage{xcolor}
\usepackage{tikz}
\usetikzlibrary{shapes.geometric, arrows}
\usetikzlibrary{decorations.markings}
\DeclareMathOperator{\im}{Im}

\newcommand{\bk}{{\bf k}}
\newcommand{\bK}{{\bf K}}
\newcommand{\bq}{{\bf q}}
\newcommand{\bQ}{{\bf Q}}

\newcommand{\bp}{{\bf p}}
\newcommand{\br}{{\bf r}}

\allowdisplaybreaks

\begin{document}
 
\title{Phonon-mediated unconventional $s$- and $f$-wave pairing superconductivity in rhombohedral stacked multilayer graphene}
\author{Emil Vi\~nas Bostr\"om}
\email{emil.bostrom@mpsd.mpg.de}
\affiliation{Max Planck Institute for the Structure and Dynamics of Matter, Luruper Chaussee 149, 22761 Hamburg, Germany}
\affiliation{Nano-Bio Spectroscopy Group, Departamento de Fisica de Materiales, Universidad del Pa\'is Vasco, 20018 San Sebastian, Spain}
\author{Ammon Fischer}
\affiliation{Institute for Theory of Statistical Physics, RWTH Aachen University, and JARA Fundamentals of Future Information Technology, 52062 Aachen, Germany}
\author{Jonas B. Hauck}
\affiliation{Institute for Theory of Statistical Physics, RWTH Aachen University, and JARA Fundamentals of Future Information Technology, 52062 Aachen, Germany}
\author{Jin Zhang}
\affiliation{Max Planck Institute for the Structure and Dynamics of Matter, Luruper Chaussee 149, 22761 Hamburg, Germany}
\author{Dante M. Kennes}
\affiliation{Institute for Theory of Statistical Physics, RWTH Aachen University, and JARA Fundamentals of Future Information Technology, 52062 Aachen, Germany}
\affiliation{Max Planck Institute for the Structure and Dynamics of Matter, Luruper Chaussee 149, 22761 Hamburg, Germany}
\author{Angel Rubio}
\email{angel.rubio@mpsd.mpg.de}
\affiliation{Max Planck Institute for the Structure and Dynamics of Matter, Luruper Chaussee 149, 22761 Hamburg, Germany}
\affiliation{Nano-Bio Spectroscopy Group, Departamento de Fisica de Materiales, Universidad del Pa\'is Vasco, 20018 San Sebastian, Spain}
\affiliation{Center for Computational Quantum Physics, Flatiron Institute, Simons Foundation, New York City, NY 10010, USA}
\date{\today}

\begin{abstract}
 Understanding the origin of superconductivity in correlated two-dimensional materials is a key step in leveraging material engineering techniques for next-generation nanoscale devices. The recent demonstration of superconductivity in Bernal bilayer and rhombohedral trilayer graphene~\cite{Zhou2021,Zhou2022}, as well as in a large family of graphene-based moir\'e systems, indicate a common superconducting mechanism across these platforms. Here we combine first principles simulations with effective low-energy theories to investigate the superconducting mechanism and pairing symmetry in rhombohedral stacked graphene multilayers. We find that a phonon-mediated attraction can quantitatively explain the main experimental findings, namely the displacement field and doping dependence of the critical temperature and the presence of two superconducting regions whose pairing symmetries depend on the parent normal state. In particular, we find that intra-valley phonon scattering favors a triplet $f$-wave pairing out of a spin and valley polarized normal state. We also propose a new and so far unexplored superconducting region at higher hole doping densities $n_h \approx 4 \times 10^{12}$ cm$^{-2}$, and demonstrate how this large hole-doped regime can be reached in heterostructures consisting of monolayer $\alpha$-RuCl$_3$ and rhombohedral trilayer graphene.
\end{abstract}

\maketitle


Recently, superconducting phases have been discovered both in Bernal bilayer graphene (BBG) and in rhombohedral trilayer graphene (RTG)~\cite{Zhou2021,Zhou2022,Pantaleon2023}. These discoveries follow an intense investigation into superconducting states of twisted bilayer and trilayer graphene, where the superconducting mechanism has been ascribed to either electron-phonon interactions or the enhanced electron-electron correlations arising in flat bands~\cite{Cao2018a,Cao2018b,Chen2019}. Similar explanations have been proposed for RTG~\cite{Ghazaryan2021,Szabo2022,Dong2021,Dong2021,Huang2022,Chou2021,Chou2022}, where a high density of states at the Fermi energy can be obtained via gate tuning in a perpendicular displacement field, leading to van Hove singularities at small but finite doping. More recently, the proximity to WSe$_2$ has been shown to increase the critical temperature of BBG by a factor ten~\cite{Chou2022b,Zhang2023}, which may be explained by the suppression of order parameter fluctuations due to an induced Ising spin-orbit coupling~\cite{Curtis2023}.

The similar phenomenology across these material platforms indicates a common mechanism underlying their superconductivity. To gain further insight into the superconducting mechanism of rhombohedral multilayer graphene, we have performed extensive first principles calculations of RTG and rhombohedral hexalayer graphene (RHG) to evaluate the phonon contribution to the superconducting pairing within Eliashberg theory. The results are in good quantitative agreement with experimental findings in RTG~\cite{Zhou2021,Zhou2022}, and predict two superconducting regions with critical temperatures $T_c \sim 100$ mK whose gap symmetries depend on the parent normal state. In particular, for a spin- and valley-polarized (SVP) parent state we find for the first time a gap with triplet $f$-wave symmetry stabilized purely through electron-phonon interactions. The quantitative improvement over previous studies of phonon-mediated pairing~\cite{Chou2021,Chou2022} can be assigned to the additional retardation effects included in the Eliashberg function, which localize the gap function to the electronic Fermi surface. We analyze the symmetry of the superconducting gap and find an extended $s$-wave pairing domain arising from inter-valley scattering and originating from a spin- and valley-unpolarized normal state. In addition, we find a smaller superconducting region with $f$-wave symmetry at lower doping levels, which is due to intra-valley scattering and arises out of an SVP normal state. This is in good agreement with the superconducting regions identified in recent experiments~\cite{Zhou2021,Zhou2022}. Compared to RTG, we find RHG shows a slightly increased critical temperature.


\begin{figure*}
 \includegraphics[width=\textwidth]{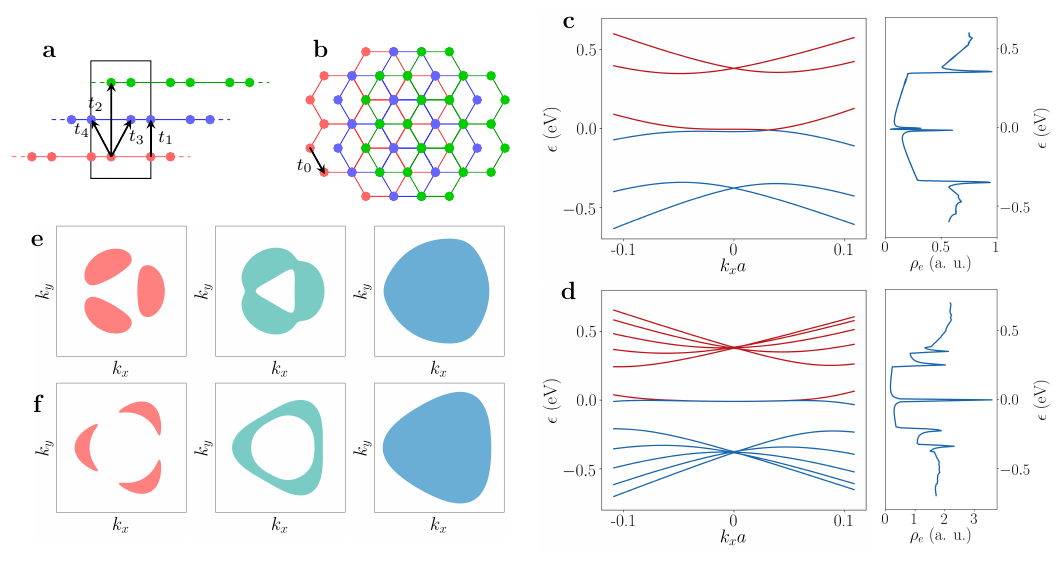}
 \caption{{\bf Electronic structure of rhombohedral stacked trilayer and hexalayer graphene.} {\bf a,} Side view of the unit cell of rhombohedral stacked trilayer and hexalayer graphene. The inter-layer hopping amplitudes $t_i$ included in the tight-binding description of the system are indicated. {\bf b,} Top view of rhombohedral stacked multi-layer graphene. The intra-layer hopping amplitude $t_0$ is indicated. {\bf c,} Low-energy band structure and density of states (DOS) of rhombohedral trilayer graphene (RTG) around the $K_+$ point. {\bf d,} Low-energy band structure and DOS of rhombohedral hexalayer graphene (RHG) around the $K_+$ point. {\bf e,} Electronic Fermi surface of RTG for zero displacement field at Fermi levels $\epsilon_F = -7.4$ meV, $\epsilon_F = -7.8$ meV and $\epsilon_F = -10.2$ meV. {\bf f,} Electronic Fermi surface of RHG for zero displacement field at Fermi levels $\epsilon_F = -3.9$ meV, $\epsilon_F = -4.5$ meV and $\epsilon_F = -10.7$ meV. The Fermi level is measured from the Dirac cone  and the DOS is calculated with a Gaussian smearing of width $\sigma = 2$ meV.}
 \label{fig:electronic_structure}
\end{figure*}

We also discover a new superconducting region in both RTG and RHG at higher hole doping densities. This region coincides with a second set of van Hove singularities further below the Dirac cone, arising from the top of the split-off valence bands. We investigate the symmetry of this new superconducting region, and again find a dominant $s$- or $f$-wave pairing depending on the spin- and valley-polarization of the parent normal state. As a mean to reach the high-doping regime we consider a heterostructure consisting of RTG and monolayer $\alpha$-RuCl$_3$, where the large work function mismatch leads to significant doping of both structures. We finally discuss the interplay of RuCl$_3$ and RTG superconductivity in the high doping regime.


\subsection*{Trilayer and hexalayer rhombohedral graphene in a finite displacement field}
The first principles electronic structure of multilayer rhombohedral stacked graphene, as obtained from density functional theory (DFT) calculations, is well-captured around the Dirac cones by a tight-binding description of the C $p_z$-orbitals including the hopping processes illustrated in Fig.~\ref{fig:electronic_structure}~\cite{Zhang2010,Shi2023}. Depending on the stacking order, the bands around the Fermi level have different characteristics: For rhombohedral stacking the highest valence band is approximately flat for $\bk \approx {\bf K}_s$, leading to a sharp van Hove singularity in the density of states (DOS). This is in contrast to Bernal stacked graphene layers, where the highest valence band approximately retains the linear dispersion of an isolated graphene sheet. The band structure and density of states (DOS) of RTG and RHG are shown in Fig.~\ref{fig:electronic_structure}.

The DOS at the van Hove singularity can further be tuned by applying a displacement field perpendicular to the graphene layers~\cite{Zhou2021} (see Supplementary Fig.~S1). To lowest order in the electronic screening, such a displacement field can be treated as a symmetric potential $\Delta$ applied across the layers. For RTG the displacement field leads to an increase in the DOS due to a gap opening at ${\bf K}_s$ and the highest valence band bending into a double-well shape. In contrast, for RHG the DOS decreases with increasing displacement field, since the highest valence band is approximately flat already at $\Delta = 0$.

\begin{figure*}
 \includegraphics[width=0.9\textwidth]{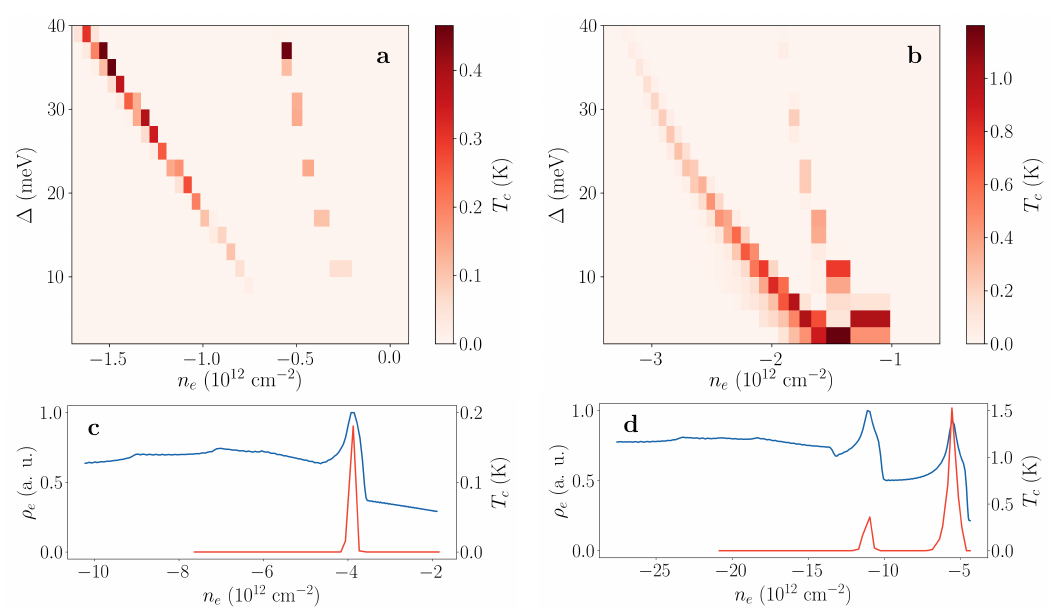}
 \caption{{\bf Superconducting critical temperature of rhombohedral tri- and hexalayer graphene.} {\bf a-b,} Superconducting critical temperature $T_c$ of rhombohedral trilayer graphene (RTG, {\bf a}) and rhombohedral hexalayer graphene (RHG, {\bf b}) as a function of electron doping density $n_e$ and displacement potential $\Delta$. {\bf c-d,} Electronic density of states (DOS, blue) and superconducting critical temperature $T_c$ (red) of RTG ({\bf c}) and RHG ({\bf d}) as a function of doping density in the high doping regime. Here $\Delta = 0$ since the bands far below the Dirac cone are insensitive to the displacement field, and the DOS is calculated with a Gaussian smearing of width $\sigma = 2$ meV.}
 \label{fig:superconductivity}
\end{figure*}


Recent experiments have found two superconducting regions in rhombohedral trilayer graphene (RTG) as a function of displacement field and doping~\cite{Zhou2021}, denoted SC1 and SC2. These superconducting phases were found to be associated with a change in the Fermi surface topology, which for finite displacement field and as a function of doping evolves from three well-separated hole-pockets, via an annular Fermi surface, to an approximately circular Fermi surface (see Fig.~\ref{fig:electronic_structure}). Similarly, the Fermi surface of RHG undergoes a transition from three isolated and strongly warped hole pockets into first an annular Fermi surface, and subsequently into a single large hole pocket. Compared to RTG the trigonal warping of the Fermi surface in RHG is more pronounced, and the total Fermi surface area larger leading to a larger DOS.


\subsection*{Eliashberg theory of phonon-mediated superconductivity}
To obtain the superconducting critical temperature in RTG and RHG resulting from phonon-mediated pairing, we use Eliashberg theory. Crucially, this approach takes into account the frequency dependence of the retarded electron-phonon interaction, which is found to have important consequences for the present systems. The key quantity of this approach is the Eliashberg function $\alpha^2 F(\omega)$, which is obtained from the spectral function of the electron-phonon self-energy $\Pi_{\nu\bq}(\omega,T)$ (see Methods for a detailed discussion). To evaluate the self-energy it is sufficient to calculate the electronic dispersion $\epsilon_{n\bk}$, the phonon frequencies $\omega_{\nu\bq}$, and the electron-phonon couplings $g_{mn\nu}(\bk,\bq)$. 

Due to the structure of the Eliashberg equations, electronic states contributing to the formation of a low-temperature superconducting instability are highly restricted to the Fermi surface. In both RTG and RHG this constitutes a very small region of the Brillouin zone (see Fig.~\ref{fig:electronic_structure}), and therefore only phonons with $\bq = 0$ or $\bq = {\bf K}_{\pm}$ contribute significantly to the superconducting pairing. The Eliashberg function can then be represented as a series of peaks, such that the effective electron-phonon coupling $\lambda$ is given by $\lambda = \sum_{\nu\bq} \lambda_{\nu\bq}$. Here $\lambda_{\nu\bq} = \gamma_{\nu\bq}/(\pi \rho_F\omega_{\nu\bq}^2)$ is the contribution to $\lambda$ from the phonon mode in branch $\nu$ and with momentum $\bq$, $\gamma_{\nu\bq}$ is the phonon linewidth, and $\rho_F$ the density of states at the Fermi level. The superconducting critical temperature follows from the McMillan equation~\cite{Allen1975}, where the average screened electron-electron interaction is accounted for by the effective parameter $\mu^{*}$. All quantities needed to evaluate $T_c$ have been calculated from first principles as discussed in the Methods.


\subsection*{Critical temperature}
First, we calculate the critical temperature $T_c$ of the superconducting transition in RTG (RHG) assuming an unpolairzed normal state as shown in Fig.~\ref{fig:superconductivity}. The results for RTG are in good quantitative agreement with the experiments of Ref.~\cite{Zhou2021}, although the critical temperature is overestimated by about a factor of four. This discrepancy is expected due to the shortcomings of Eliashberg theory to incorporate the effect of strong quantum fluctuations in two-dimensional systems, which tend to suppress the critical temperature. Due to the higher DOS of RHG as compared to RTG, the critical temperature in this system is enhanced by a factor $2-3$. To account for the superconducting region SC2 arising out of an SVP state, we assume that the effective DOS in the active channel is increased by a factor of four (since all electrons are accumulated in one spin and valley sector), and that inter-valley scattering is negligible. This gives a superconducting region at lower doping with slightly lower $T_c$ compared to the phase SC1, in good agreement with experiment.

For both RTG and RHG the critical temperature is found to closely follow the electronic DOS, which is a consequence of the self-energy scaling like $\Pi \sim \rho_F^2$, such that $\alpha^2 F(\omega) \sim \rho_F$. Since the DOS increases (decreases) with increasing displacement field in RTG (RHG) (see Supplementary Fig.~S1), the critical temperatures show the same trends. We note that compared to conventional BCS theory~\cite{Chou2021,Chou2022}, the main consequence of including phonon retardation effects is an overall reduction of the critical temperature by a factor $5 - 10$ and the range of dopings over which superconductivity is found.

The DOS and critical temperature in the high doping regime are also shown in Fig.~\ref{fig:superconductivity}. The density of states is similar to that found in the low doping regime, and the corresponding critical temperature is therefore of the same order. Since the active bands in the high doping regime arise from the intermediate layers of the RTG and RHG stacks, their energies are largely insensitive to the displacement field.


\subsection*{Linearized gap equation}
Next, we analyze the symmetry of the leading superconducting order parameter in both doping regimes. The gap function can be obtained by solving the linearized gap equation~\cite{Allen1983,Giustino2017}
\begin{align}
 \Delta_{m\bk} &= \pi \sum_{n\bp} \chi_{mn\bk\bp} \frac{\tanh(\beta \xi_{n\bp})}{\beta \xi_{n\bp}} \Delta_{n\bp},
\end{align}
where $\beta$ is the inverse temperature, $\xi_{n\bk} = \epsilon_{n\bk} - \epsilon_F$ and the susceptibility is 
\begin{align}
 \chi_{mn\bk\bp} &= \sum_\nu \frac{2|g_{mn\nu}(\bk,\bq)|^2}{N_\bk\rho_F \omega_{\bq\nu}} \delta(\xi_{m\bk}) \delta(\xi_{n\bp}).
\end{align}
Here it is implicitly assumed that the phonon momentum satisfies $\bq = \bp - \bk$. This equation is of the same form as the standard BCS gap equation, however with the interaction restricted to the Fermi surface. The main difference between BCS and Eliashberg theory is that the latter includes retardation effects from the electron-phonon interaction, which give rise to a non-trivial frequency dependence in the effective electron-electron interaction (see Methods). This frequency dependence significantly improves the temperature dependence of the theory, and in the low temperature limit localizes the susceptibility to the Fermi surface~\cite{Allen1983,Giustino2017}. For many systems, where the Fermi surface occupies a significant portion of the Brillouin zone, this effect is rather small. For RTG and RHG however, where the Fermi surface occupies a tiny portion of the full Brillouin zone, the quantitative difference between the two approaches is quite dramatic. In general Eliashberg theory is expected to be more accurate than BCS theory, since it captures the dynamical aspects of the electron-phonon interaction.


\begin{figure*}
 \includegraphics[width=\textwidth]{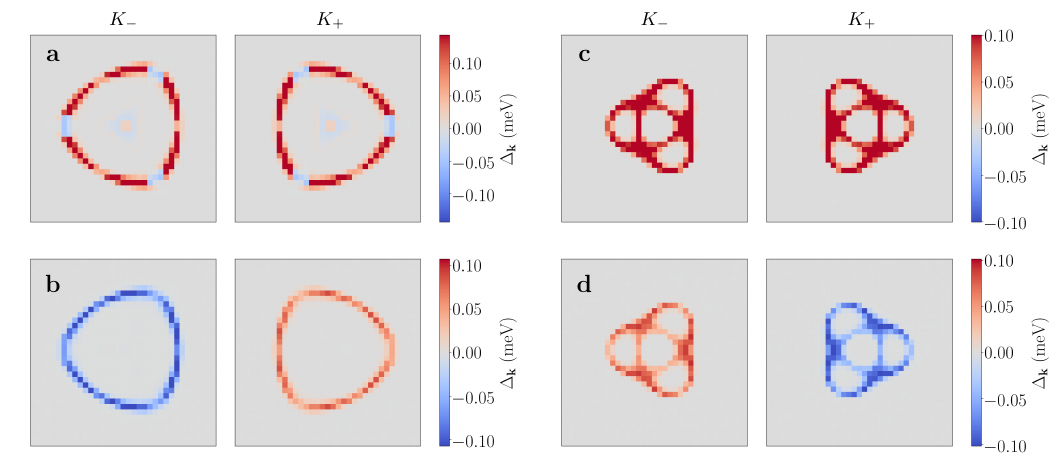}
 \caption{{\bf Superconducting gap of rhombohedral trilayer graphene.} {\bf a-b,} Superconducting gap $\Delta_\bk$ of rhombohedral stacked trilayer graphene for a hole doping density $n_h \approx 0.4 \times 10^{12}$ cm$^{-2}$, including the attractive interaction from both inter- and intra-valley phonon scattering ({\bf a}) or only intra-valley scattering ({\bf b}). The dominant inter-valley scattering favors $s$-wave pairing ({\bf a}), while the sub-dominant intra-valley scattering favors $f$-wave pairing ({\bf b}). {\bf c-d,} Superconducting gap $\Delta_\bk$ of rhombohedral stacked trilayer graphene for a hole doping density $n_h \approx 3.9 \times 10^{12}$ cm$^{-2}$, including the attractive interaction from both inter- and intra-valley phonon scattering ({\bf c}), or only intra-valley scattering ({\bf d}). The dominant inter-valley scattering favors $s$-wave pairing ({\bf c}), while the sub-dominant intra-valley scattering favors $f$-wave pairing ({\bf d}). In all panels the momentum runs over a region $ka \in [-0.1, 0.1]$ around the $K_s$ point.}
 \label{fig:superconducting_gap}
\end{figure*}


\subsection*{Superconducting symmetry in the low-doping regime}
Fig.~\ref{fig:superconducting_gap} shows the superconducting gap of RTG for a displacement field potential $\Delta = 20$ meV and the Fermi level at $\epsilon_F = -26$ meV, corresponding to a hole doping density of $n_h \approx 0.4 \times 10^{12}$ cm$^{-2}$. As can be clearly seen the gap is symmetric under inversion, $\Delta_{-\bk} = \Delta_\bk$, indicating a singlet pairing. Further, in each valley the gap has three-fold rotational symmetry with a non-trivial nodal structure. Together these observations are consistent with an extended $s$-wave symmetry. A similar gap structure is found for RHG, again consistent with an extended $s$-wave symmetry (see Supplementary Fig.~S2). This dominant pairing is found to arise from inter-valley scattering, corresponding to the exchange of virtual phonons with momenta $\bq \approx \bK$, as expected from a symmetry analysis (see Methods for an extended discussion).

Artificially suppressing the inter-valley scattering, as appropriate for a superconducting state arising out of an SVP normal state, we find an odd gap $\Delta_{-\bk} = -\Delta_\bk$. This indicates an unconventional triplet $f$-wave pairing, and that superconductivity in RTG arises from a competition between inter- and intra-valley scatterings favoring different superconducting symmetries. Again, the results for RHG are qualitatively similar (see Supplementary Fig.~S2). These findings are in good agreement with the experimental results of Ref.~\cite{Zhou2021}, where the superconducting region SC2 was found to violate the Pauli limit by more than an order of magnitude, strongly indicating unconventional superconducting pairing.

The $f$-wave pairing is likely stabilized by a combination of Fermi surface topology and the spin and valley polarization of the parent state. In fact, for an unpolarized parent state with momentum-independent electron-phonon scattering, the $s$- and $f$-wave pairings are found to be degenerate (see Supplemental Material). This indicates a highly non-trivial interplay of electronic correlations and phonon-mediated pairing, where the former stabilizes the parent state and the latter the $f$-wave pairing.


\subsection*{Superconducting symmetry in the high-doping regime}
Similar results are found for both RTG and RHG in the high-doping regime, where inter-valley (intra-valley) scattering is found to favor an extended $s$-wave ($f$-wave) pairing. The superconducting gap for RTG at zero displacement field and a Fermi level of $\epsilon_F = -350$ meV, corresponding to a hole doping density of $n_h \approx 3.9 \times 10^{12}$ cm$^{-2}$, is shown in Fig.~\ref{fig:superconducting_gap}. We note that the critical temperature in the high doping regime is comparable but slightly lower than that of the low doping regime, as expected from the respective DOS. These results indicate a new and so far unexplored region of superconductivity in RTG, arising from the van Hove singularities of the lower valence bands.


\subsection*{Reaching the high-doping regime}
To reach the high doping regime, we consider a heterostructure consisting of RTG and a monolayer of the Mott-Slater insulator $\alpha$-RuCl$_3$. This heterostructure has recently been found to realize a heavily hole-doped regime of graphene, with a Fermi level $\sim 0.6$ eV below the Dirac cone~\cite{Rizzo2020}. More specifically, due to the large work function mismatch of about $1.6$ eV, there is a significant charge transfer from the graphene multilayer into $\alpha$-RuCl$_3$ resulting in an overall doping of approximately $-0.07e$ per Ru atom~\cite{Biswas2019,Rizzo2020}. Since the charge transfer is mainly localized to the interface, the charge transfer corresponds to a hole doping of $0.01e$ per C atom, or equivalently a doping density of $4.2 \times 10^{12}$ cm$^{-2}$, in the layer adjacent to $\alpha$-RuCl$_3$. 

\begin{figure}
 \includegraphics[width=\columnwidth]{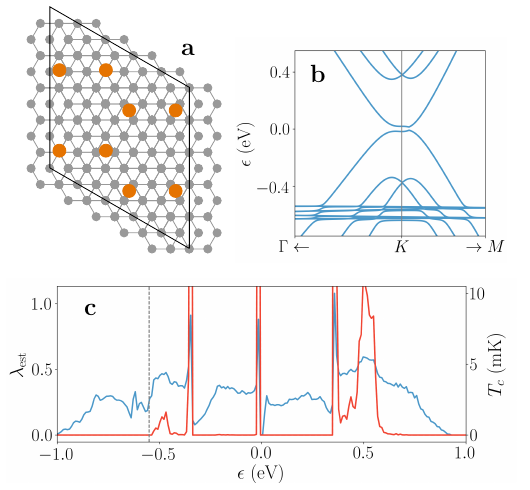}
 \caption{{\bf Superconductivity of a rhombohedral trilayer graphene and $\alpha$-RuCl$_3$ heterostructure.} {\bf a,} Supercell of the rhombohedral trilayer graphene (RTG) and monolayer $\alpha$-RuCl$_3$ heterostructure considered, containing $8$ Ru atoms (orange) and $150$ C atoms (gray). {\bf b,} Electronic band structure of the heterostructure, with the flat bands originate from the Ru $t_{2g}$ orbitals and the dispersive bands from the C $p_z$ orbitals. {\bf c,} Estimated electron-phonon coupling $\lambda_{\rm est}$ (blue) for the electron-phonon couplings $g_{\rm RTG} = 200$ meV and $g_{\rm RuCl_3} = 20$ meV, and a phonon frequency $\omega = 200$ meV, as well as the corresponding critical temperature $T_c$. The gray dashed line indicates the Fermi level of the heterostructure.}
 \label{fig:heterostructure}
\end{figure}

The band structure of the heterostructure was calculated in a $5\times 5$ and $2\times 2$ supercell for the RTG and $\alpha$-RuCl$_3$ sub-systems respectively, is shown in Fig.~\ref{fig:heterostructure}. The local interactions in the active Ru manifold are described by a Hubbard-Kanamori Hamiltonian~\cite{Winter16,Bostrom2022} treated within the unrestricted Hartree-Fock approximation. The interaction between the subsystems is treated as a position and orbital dependent hybridization~\cite{Shi2023}, which largely restricts the coupling to the graphene layer adjacent to $\alpha$-RuCl$_3$ (see Supplemental Material). We note that the band structure displays clear avoided crossings at the band intersections, and that a gap opens at the Dirac cone of the RTG bands. This effect is in agreement with first principles calculations (see Supplementary Fig.~S3), indicating that the heterostructure sets up an intrinsic displacement field through the charge transfer process.

\subsection*{Superconductivity of the RTG-RuCl$_3$ heterostructure}
We now use Eliashberg theory to obtain the critical temperature of the heterostructure. To estimate $T_c$ we calculate the dimensionless variable $\lambda_{\rm est} = g^2 \rho/\Omega$ (see Fig.~\ref{fig:heterostructure}), which is closely related to the dimensionless electron-phonon coupling $\lambda$. The electron-phonon couplings of the separate subsystems, obtained from first principles density functional perturbation theory calculations, are found to be on the order of $g_{\rm RTG} \approx 200$ meV~\cite{Piscanec2004} and $g_{\rm RuCl_3} \approx 20 - 40$ meV, and assuming a typical optical phonon energy of $\Omega = 200$ meV as appropriate for graphene, we find $\lambda_{\rm est} \approx 1$ around the graphene van Hove singularities. This estimate is in good agreement with the more detailed calculations for RTG based on the Eliashberg function. The resulting critical temperature agrees well with Fig.~\ref{fig:superconductivity} around the van Hove singularities, and is on the order of a few mK close to the heterostructure Fermi level.

Depending on the electron-phonon coupling in $\alpha$-RuCl$_3$ as well as on the sub-system hybridization strength, the effective coupling $\lambda_{\rm est}$ of the heterostructure can be modified by a few percent. However, while a stronger electron-phonon coupling in $\alpha$-RuCl$_3$ is found to enhance $\lambda_{\rm est}$, a stronger hybridization predominantly reduces the effective electron-phonon coupling in RTG dominated bands. This effect can be attributed to the much smaller Fermi surface of RTG as compared to $\alpha$-RuCl$_3$, which makes the contribution to $\lambda_{\rm est}$ from RTG much more sensitive to the hybridization than the contribution from $\alpha$-RuCl$_3$. A small enhancement of $T_c$ with increasing hybridization can be observed for bands with a dominant projection on $\alpha$-RuCl$_3$.


\section*{Discussion}
By combining first principles calculations with effective low-energy models, we have investigated the phenomenology of phonon-mediated superconductivity in RTG, RHG and RTG-$\alpha$-RuCl$_3$ heterostructures. Including retardation effects of the phonon-mediated attraction we find a substantial reduction of both the critical temperature and the region of doping and displacement fields in which superconductivity appears. These effects lead to a substantial improvement between theory and experiments~\cite{Zhou2021,Zhou2022}, promoting phonon mediated superconductivity as a strong contender to explain superconductivity across a wide range of graphene platforms~\cite{Zhou2021,Zhou2022,Cao2018a,Cao2018b,Chen2019}.

More surprisingly, we find that inter-valley and intra-valley phonon scattering favors superconductivity with different symmetry, such that phonon mediated pairing can stabilize both extended $s$-wave as well as unconventional $f$-wave triplet pairing. More specifically, we find that $s$-wave pairing is dominant when superconductivity arises out of an unpolarized parent state, while $f$-wave pairing dominates when the parent state is SVP. If verified, this would be the first time phonons are found to stabilize unconventional triplet superconductivity~\cite{Schrodi2023}. Further, the interplay of electronic correlations and phonon-mediated pairing demonstrated by the $f$-wave pairing, indicates a path to realize unconventional triplet pairing in conventional phonon-mediated extended $s$-wave superconductors.

We finally demonstrate that a new and so far unexplored superconducting region, stemming from the van Hove singularity of the lower valence bands, can be reached in heterostructures of RTG and monolayer $\alpha$-RuCl$_3$. Although a strong inter-layer hybridization is found to suppress superconductivity in the supercell considered here, it is possible that maximizing the Fermi surface overlaps of the two subsystems~\cite{Shi2023}, e.g. via layer twisting, could lead to an enhancement of $T_c$. It might also be the case that reducing spin fluctuations in the heterostructure by breaking the spin rotational symmetry could lead to an enhancement of the critical temperature for spin triplet pairings, as have recently been proposed for Bernal bilayer graphene in proximity to WSe$_2$~\cite{Zhou2022,Curtis2023}. In fact, our first principle calculations show a small induced Ising spin-orbit coupling of $\sim 1$ meV in the heterostructure (see Supplementary Fig.~S3).

Further work is needed to conclusively determine the superconducting mechanism in RTG and RHG, taking into account electron-phonon as well as electron-electron interactions on an equal footing. In any case our work shows that the exotic and unconventional superconducting phenomenology of RTG is consistent with a phonon-mediated pairing.


\bibliography{references}


\clearpage
\section*{Methods}

\subsection*{Electronic structure, phonon dispersion and electron-phonon coupling}
The electronic structure of rhombohedral and Bernal stacked trilayer and hexalayer graphene were calculated as a function of displacement field using the {\sc Abinit} electronic structure code. The electronic ground state was calculated within the local density approximation (LDA) on a $72 \times 72$ $\bk$-point grid, using a plane wave cut-off of $20$ Ha. The displacement field was included via a coupling to the polarization as described within the modern theory of polarization~\cite{Vanderbilt2018}. The Kohn-Sham Bloch functions were subsequently transformed to maximally localized Wannier functions using the {\sc Wannier90} code, including the lowest lying 15 (30) bands for trilayer (hexalayer) systems. The Wannier functions were used to interpolate the electronic band structure to arbitrary $\bk$-points, allowing to calculate the density of state (DOS) on a very dense $\bk$-point grid with $1200 \times 1200$ grid points.

The phonon band structure was obtained using density functional perturbation theory (DFPT) as implemented in the {\sc Abinit} electronic structure code. The atomic structure was first relaxed to obtain maximal forces below $5 \cdot 10^{-6}$ Ha/Bohr, after which the phonon frequencies where calculated on a $12 \times 12$ $\bq$-point grid. To calculate the electron-phonon coupling the electronic bands were interpolated using star functions~\cite{Pickett1988}, allowing to densely sample the region of the Brillouin zone around the $K^\pm$ points (with a density corresponding to a $720 \times 720$ $\bk$-point grid). Similarly, the derivatives of the Kohn-Sham potential were Fourier interpolated to obtain the electron-phonon coupling on an arbitrarily dense $\bq$-point grid, here taken to consist of $72 \times 72$ points.


\subsection*{Eliashberg theory of phonon-mediated superconductivity}
To obtain the superconducting critical temperature in RTG and RHG resulting from phonon-mediated pairing, we use Eliashberg theory. Crucially, this approach takes into account the frequency dependence of the retarded electron-phonon interaction, which is found to have important consequences for the current systems. The key quantity of this approach is the Eliashberg function $\alpha^2 F(\omega)$, defined as
\begin{align}
 \alpha^2 F(\omega) = \frac{1}{N_q \rho_F} \sum_{\nu\bq} \frac{\gamma_{\nu\bq}}{\omega_{\nu\bq}} \delta(\omega - \omega_{\nu\bq})
\end{align}
where $N_q$ is the number of $\bq$-points for the phonons, $\rho_F$ the electronic density of states (per spin) at the Fermi energy, and $\gamma_{\nu\bq}$ and $\omega_{\nu\bq}$ is the linewidth and frequency of phonon mode $(\nu\bq)$. The phonon linewidth is related to the phonon self-energy by $\gamma_{\nu\bq} = 2\im \Pi_{\nu\bq}(\omega_{\nu\bq},T)$, where the self-energy results from electron-phonon scattering is given by
\begin{align}\label{eq:self_energy}
 \Pi_{\nu\bq}(\omega,T) = \frac{2}{N_\bk} \sum_{\bk mn} \frac{|g_{mn\nu}(\bk,\bq)|^2 (f_{n\bk} - f_{m,\bk+\bq})}{\epsilon_{n\bk} - \epsilon_{m,\bk+\bq}-\omega-i\eta}.
\end{align}
To evaluate the self-energy it is sufficient to calculate the electronic dispersion $\epsilon_{n\bk}$, the phonon frequencies $\omega_{\nu\bq}$, and the electron-phonon couplings $g_{mn\nu}(\bk,\bq)$. 

Due to the presence of the Fermi functions in Eq.~\ref{eq:self_energy}, the sum over momenta is at low temperatures highly restricted to the electronic Fermi surface. In both RTG and RHG this constitutes a very small region of the Brillouin zone (on the order of $10^{-3}G$, see Fig.~\ref{fig:electronic_structure}), and therefore only phonons with $\bq = 0$ or $\bq = \bQ = {\bf K}_+ - {\bf K}_-$ contribute significantly to the superconducting pairing. The Eliashberg function can then be represented as a series of peaks, such that the effective electron-phonon coupling $\lambda$ is given by
\begin{align}\label{eq:lambda}
 \lambda = \int \frac{\alpha^2F(\omega)}{\omega} d\omega = \sum_{\nu\bq} \lambda_{\nu\bq}.
\end{align}
Here $\lambda_{\nu\bq} = \gamma_{\nu\bq}/(\pi \rho_F\omega_{\nu\bq}^2)$ is the contribution to $\lambda$ from mode $(\nu\bq)$. The superconducting critical temperature can now be obtained from the McMillan equation~\cite{Allen1975}
\begin{align}\label{eq:tc}
 T_c = \frac{\omega_{\rm log}}{1.2} \exp\left[ \frac{-1.04 (1+\lambda)}{\lambda(1-0.62\mu^{*}) - \mu{*}} \right],
\end{align}
where $\omega_{\rm log}$ is a logarithmic average of the phonon frequencies, and $\mu^{*}$ is the average screened electron-electron interaction strength (here taken to have the typical value $\mu^{*} = 0.2$). All the quantities needed to evaluate $T_c$ have been calculated from first principles as discussed below.


\subsection*{Gap equation}
Within Eliashberg theory, the gap equation is of the form~\cite{Allen1983,Giustino2017}
\begin{align}
 &Z_{n\bk}(i\omega_m) = 1 + \pi k_B T \sum_{m'n'\bk'} \frac{(\omega_{m'}/\omega_m)}{\sqrt{\omega_{m'}^2 + \Delta_{n'\bk'}^2(i\omega_{m'})}} \nonumber \\
 &\hspace*{1cm}\times \lambda_{n\bk,n'\bk'}(i\omega_m - i\omega_{m'}) \\
 &Z_{n\bk}(i\omega_m) \Delta_{n\bk}(\omega_m) = \pi k_B T \sum_{m'n'\bk'} \frac{\Delta_{n'\bk'}}{\sqrt{\omega_{m'}^2 + \Delta_{n'\bk'}^2(i\omega_{m'})}} \nonumber \\
 &\hspace*{1cm}\times [ \lambda_{\bk\bk'}^{nn'}(i\omega_m - i\omega_{m'}) - N_F U_{\bk\bk'}^{nn'}]
\end{align}
where $U_{\bk\bk'}^{nn'}$ is the screened Coulomb interaction between momenta $\bk$ and $\bk'$ on the Fermi surface, and the electron-phonon coupling is
\begin{align}
  \lambda_{\bk\bk'}^{nn'}(i\omega) &= N_F^{-1} \sum_\nu \frac{2\omega_{\bq\nu}}{\omega_{\bq\nu}^2 + \omega^2} |g_{nn'}(\bk,\bq)|^2 \\
 &\times \delta(\epsilon_\bk - \epsilon_F) \delta(\epsilon_\bk' - \epsilon_F). \nonumber
\end{align}
In this equation the phonon momentum has to satisfy $\bq = \bk' - \bk$. Assuming $Z \approx 1$ and taking the static limit, we find the gap equation
\begin{align}
 &\Delta_{n\bk} = \pi k_B T \sum_{in'\bk'} \frac{\Delta_{n'\bk'}}{\sqrt{\omega_{i}^2 + \Delta_{n'\bk'}^2}} [ \lambda_{\bk\bk'}^{nn'} - U_{\bk\bk'}^{nn'}] \\
 &\lambda_{n\bk,n'\bk'}(i\omega) = \sum_\nu \frac{2|g_{nn'}(\bk,\bq)|^2}{N_F N_\bk \omega_{\bq\nu}} \delta(\epsilon_\bk - \epsilon_F) \delta(\epsilon_{\bk'} - \epsilon_F). \nonumber
\end{align}
Linearizing this equation we find an equation of the same form as the BCS gap equation, 
\begin{align}
 \Delta_{n\bk} &= \pi \sum_{n'\bk'} \lambda_{n\bk,n'\bk'} \frac{\tanh(\beta \xi_{n'\bk'})}{\beta \xi_{n'\bk'}} \Delta_{n'\bk'},
\end{align}
although with the interaction restricted to the Fermi surface.

As a check on this result, the effective electron-phonon coupling can be compared to that derived from the electron-phonon self-energy. Recalling that the phonon linewidth is
\begin{align}
 \gamma_{mn\nu}^\bq(T) &= -\frac{2\omega_{\bq\nu}}{N_\bk} \sum_{\bk\bk' mn} |g_{mn\nu}(\bk,\bq)|^2 \delta(\epsilon_{\bk n} - \epsilon_F) \\
 &\times \delta(\epsilon_{\bk'm} - \epsilon_F). \nonumber
\end{align}
This results in the electron-phonon coupling
\begin{align}
  \lambda_{n\bk\,n'\bk'} &= \sum_{\nu} \frac{2 |g_{nn'}(\bk,\bq)|^2}{\pi N_F N_\bk \omega_{\bq\nu}} \delta(\epsilon_{\bk} - \epsilon_F) \delta(\epsilon_{\bk'} - \epsilon_F)
\end{align}
From these considerations, we note that the approximate Eliashberg gap equation is obtained from the BCS gap equation by including a factor $N_F^{-1} \delta(\epsilon_{\bk n} - \epsilon_F) \delta(\epsilon_{\bk' n'} - \epsilon_F)$ in the susceptibility. The main difference between these approaches is therefore that the Eliashberg treatment localizes the susceptibility to the Fermi surface. For most metallic systems, where the Fermi surface occupies a significant portion of the Brillouin zone, this difference might not be so severe. For multilayer graphene systems however, where the Fermi surface is a tiny portion of the full Brillouin zone, the difference between the approaches is quite dramatic.


\subsection*{Gap symmetry analysis}
The symmetry of the gap is determined by the point group of the material, and rhombohedral multilayer graphene belongs to the point group $D_{3d}$. The wave functions at $\bK_+$ and $\bK_-$ therefore have to satisfy the symmetry constraint $C_3 \psi(\br) C_3^{-1} = e^{(2\pi i/3) \tau_z \sigma_z} \psi(R_3\br)$, where $\tau_z$ ($\sigma_z$) is a Pauli matrix in valley (sublattice) space. In both RTG and RHG, intra-valley pairing is associated with finite momentum Cooper pairs and will therefore be strongly suppressed. This follows from the inequivalence $\epsilon_{\tau n}(\bk) \neq \epsilon_{\tau n} (-\bk)$, where $\tau$ and $n$ are valley and band indexes, such that no nesting conditions are met (see Fig.~\ref{fig:electronic_structure}). It is therefore expected that inter-valley pairing is the dominant mechanism in these systems. 

For inter-valley Cooper pairs, the symmetry constraint implies that intra-sublattice pairings are invariant under $C_3$ symmetry, while inter-sublattice pairings will acquire a net phase. The former property is expected of $s$- and $f$-wave pairings ($C_3 \Delta_\bk C_3^{-1} = \Delta_\bk$), while the later is expected for $p$- and $d$-wave pairings ($C_3 \Delta_\bk C_3^{-1} = e^{i\phi} \Delta_\bk$). Therefore, $s$- and $f$-wave symmetries can be distinguished from $p$- and $d$-wave symmetries by looking at how the gap transforms under three-fold rotations. Similarly, $s$-wave symmetry can be distinguished from $f$-wave symmetry by the behavior under inversion, since $s$- and $d$-wave gaps have to be singlets ($\Delta_{-\bk} = \Delta_\bk$), while $p$- and $f$-waves gaps have to be triplets ($\Delta_{-\bk} = -\Delta_\bk$).

For the low-energy bands of RTG and RHG, the wave function is mainly localized to the dangling sites $A_1$ and $B_N$ (with $N$ the number of layers), such that inter-sublattice pairings are strongly suppressed. Therefore, the dominant pairing in these bands is expected to be either $s$- or $f$-wave. However, for the lower valence bands at $\sim -0.5$ eV, other pairing channels might become competitive through inter-sublattice scattering in the intermediate layers.


\section*{Acknowledgements} 
We acknowledge support from the Cluster of Excellence ``CUI: Advanced Imaging of Matter''- EXC 2056 - project ID 390715994 and SFB-925 ``Light induced dynamics and control of correlated quantum systems'' – project 170620586 of the Deutsche Forschungsgemeinschaft (DFG), and Grupos Consolidados (IT1453-22). EVB acknowledges funding from the European Union's Horizon Europe research and innovation programme under the Marie Sk{\l}odowska-Curie grant agreement No 101106809. We further acknowledge support from the Max Planck-New York City Center for
Non-Equilibrium Quantum Phenomena. The Flatiron Institute is a division of the Simons Foundation.

\section*{Data Availability}
All data supporting the findings of this study are available from the corresponding authors upon reasonable request.

\section*{Code Availability}
The codes used to generate the data of this study are available from the corresponding authors upon reasonable request.

\section*{Competing Interests}
The authors declare no competing financial or non-financial interests.

\section*{Author Contributions}
EVB performed the first principles calculations of RTG, RHG and monolayer $\alpha$-RuCl$_3$, and JZ performed the first principles calculations of the RTG-$\alpha$-RuCl$_3$ heterostructure. EVB performed the Eliashberg theory calculations of the superconducting properties with input from AF, JH, DK and AR. DK and AR supervised the work, and all authors collaborated in writing the manuscript.


\end{document}